\documentclass[12pt]{article}
\usepackage{graphicx}
\usepackage{enumerate}
\usepackage{cite}
\def\lsim{\mathrel{\vcenter{\hbox{$<$}\nointerlineskip\hbox{$\sim$}}}}
\newcommand{\be}{\begin{equation}}
\newcommand{\ee}{\end{equation}}
\newcommand{\ba}{\begin{eqnarray}}
\newcommand{\ea}{\end{eqnarray}}

\def\21{$SU(2) \otimes U(1) $}

\def\lsim{\raise0.3ex\hbox{$\;<$\kern-0.75em\raise-1.1ex\hbox{$\sim\;$}}}
\def\gsim{\raise0.3ex\hbox{$\;>$\kern-0.75em\raise-1.1ex\hbox{$\sim\;$}}}

\newcommand{\mx}{\left[\begin{array}}
\newcommand{\finmx}{\end{array}\right]}
\newcommand{\mxp}{\left(\begin{array}}
\newcommand{\finmxp}{\end{array}\right)}
\def\beq{\begin{equation}}
\def\eeq{\end{equation}}
\def\bea{\begin{eqnarray}}
\def\eea{\end{eqnarray}}

\def\mathbf#1{\hbox{\bf #1}}
\def\textrm#1{\hbox{#1}}

\def\lsim{\raise0.3ex\hbox{$\;<$\kern-0.75em\raise-1.1ex\hbox{$\sim\;$}}}
\def\gsim{\raise0.3ex\hbox{$\;>$\kern-0.75em\raise-1.1ex\hbox{$\sim\;$}}}
\newcommand {\ignore}[1]{}

\begin{document}
\vspace*{-1in}
\renewcommand{\thefootnote}{\fnsymbol{footnote}}
\begin{flushright}
\texttt{
}
\end{flushright}
\vskip 5pt
\begin{center}
{\Large{\bf Double beta decay versus cosmology:
Majorana CP phases and nuclear matrix elements}}
\vskip 25pt

{\sf
Frank Deppisch$^{1}$\footnote[1]{E-mail: deppisch@physik.uni-wuerzburg.de},
Heinrich P\"as$^1$\footnote[2]{E-mail: paes@physik.uni-wuerzburg.de},
Jouni Suhonen $^{2}$\footnote[1]{E-mail: jouni.suhonen@phys.jyu.fi}
\vskip 10pt
{\it \small $^1$ Institut f\"ur Theoretische Physik und Astrophysik,\\
Universit\"at W\"urzburg, D-97074 W\"urzburg, Germany}\\
{\it \small
$^2$ Department of Physics, University of Jyv\"askyl\"a, \\P.O.B. 35, 
FIN-40014, Jyv\"askyl\"a, Finland}}

\vskip 20pt

{\bf Abstract}
\end{center}

\begin{quotation}
{\small\noindent
We discuss the relation between the absolute neutrino mass scale, the
effective mass measured in neutrinoless double beta decay, and the
Majorana CP phases. Emphasis is placed on estimating the upper bound on
the nuclear matrix element entering calculations of the double beta decay
half life. Consequently,
one of the Majorana CP phases can be constrained when combining
the claimed evidence for neutrinoless double beta decay with the neutrino
mass bound from cosmology.
}
\end{quotation}

\vskip 20pt

\setcounter{footnote}{0}
\renewcommand{\thefootnote}{\arabic{footnote}}

\newpage

Over the past years much effort has been invested to probe
leptonic mixing with increasing accuracy, and, in fact, a unique
picture is evolving from the precise measurements of neutrino
oscillation probabilities. For a full construction of the mixing
matrix, however, knowledge about CP violating phases is necessary.

To describe leptonic mixing one can always work in a basis where
the charged lepton Yukawa matrix is diagonal. In this case the
neutrino mass matrix $m^{\nu}$ can be written in the flavor basis
as
\be\label{eqn:mnu}
m^{\nu}=U^* m^{\rm diag} U^{\dagger},
\ee
with $m^{\rm diag}={\rm diag}(m_1,m_2,m_3)$, and the
Maki-Nakagawa-Sakata (MNS) matrix \(U\), which contains three
mixing angles and one CP violating Dirac phase. The determination
of the mixing angles is subject to neutrino oscillation
experiments, as is (at least in principle) the determination of
the Dirac phase - the leptonic analogue of the CKM phase in the
quark sector. Three of the mixing angles can be identified with
the maximal, large and small observables measured in atmospheric,
solar and reactor neutrino oscillations, respectively \cite{Maltoni:2004ei}. 
Important information on the Dirac phase can be expected from
a combination of future long-baseline experiments
\cite{Huber:2004ug}, and ultimately, from
a neutrino factory \cite{Gomez-Cadenas:2001ev}.

If the neutrino is of Majorana type, two more phases enter,
though. The determination of these Majorana phases is, in fact, the
most challenging task in the reconstruction of the fundamental
parameters of the Standard Model particle content. In general,
\(U\) can be written as
\be{}
U=V\cdot {\rm diag}(1, e^{i
\phi_{12}/2},e^{i \phi_{23}/2}),
\ee
where $V$ is parametrized in
the standard CKM form, \beq
  V=\left( \begin{array}{ccc} c_{13}c_{12}          & c_{13}s_{12}           & s_{13}e^{-i\delta}    \\
-c_{23}s_{12}-s_{23}s_{13}c_{12}e^{i\delta} & c_{23}c_{12}-s_{23}s_{13}s_{12}e^{i\delta} & s_{23}c_{13} \\
s_{23}s_{12}-c_{23}s_{13}c_{12}e^{i\delta} &-s_{23}c_{12}-c_{23}s_{13}s_{12}e^{i\delta} & c_{23}c_{13}
\end{array} \right),
\eeq
and $\phi_{ij}$ are the Majorana phases under discussion.
A viable possibility to obtain information on the Majorana phases
is to compare measurements of the absolute neutrino masses $m_i$
with the elements of the neutrino mass matrix $m^{\nu}$ in the flavor basis.
While absolute neutrino masses are most stringently constrained from cosmology,
only the $ee$ element of $m^{\nu}$ is experimentally accessible, being the
effective mass $m_{ee}$ measured in
neutrinoless double beta decay. Several works have discussed the
relations of $m_{ee}$, absolute neutrino masses and Majorana phases
\cite{phasestheory}.
In
\cite{Matsuda:2003kf} it was pointed out that one could restrict the
Majorana Phase $\phi_{12}$ by using the recently claimed evidence for
neutrinoless double beta decay \cite{HMexp}
and the cosmological neutrino mass bound derived by the
WMAP collaboration \cite{wmap},
if a certain nuclear matrix element (NME) calculation \cite{smb} is assumed.
An important issue is, however, the uncertainty in the neutrino mass 
determination within the double beta decay framework
due to systematical limitations in such NME calculations.
In this work we focus on what can be learned about Majorana phases
from recent double beta decay and cosmological structure formation data,
in view of an upper bound on the NME.
In the following, we first review the claimed evidence for neutrinoless double
beta decay and discuss the upper bound on the NME. Finally, we compare
the lower bounds on $m_{ee}$ obtained with
the upper bounds on neutrino masses from cosmology.

The half life of neutrinoless double beta decay is given by
\be{}
[T^{0\nu\beta\beta}_{1/2}]^{-1}=\left|\frac{m_{ee}}{m_e}\right|^2 G_1^{(0\nu)}
\left|{\cal M}^{(0\nu)}\right|^2,
\ee
where $m_e$ denotes the electron rest
mass, $G_1^{(0\nu)}$ is a phase space factor,
and the NME is given by
${\cal M}^{(0\nu)}={\cal M}_{GT}-{\cal M}_{F}$, being a  combination of 
Gamov-Teller and Fermi transitions.

The neutrinoless double beta decay is sensitive to the $ee$ element of
the mass matrix $m^{\nu}$ (\ref{eqn:mnu}) in flavor space,
\ba
|m_{ee}| &=& \left|\sum_i |V_{ei}|^2 e^{i \phi_i} m_i\right| \nonumber \\
&=& \left|m_1 |V_{e1}|^2
       + \sqrt{m^2_1 + \Delta m_{12}^2}  |V_{e2}|^2 e^{i \phi_{12} }
       + \sqrt{m^2_1 + \Delta m_{12}^2 \pm \Delta m_{23}^2}  
       |V_{e3}|^2 e^{i \phi_{23}}\right|.\label{mee}
\ea
Here the mass eigenstates are expressed as $m_1$, $m_2 = \sqrt{m_1^2 
+ \Delta m_{12}^2}$,
$m_3= \sqrt{m_2^2 \pm \Delta m_{23}^2}$, where the plus 
(minus) sign
applies in the normal (inverted) mass hierarchy case.


From the Chooz and Palo Verde experiments we know that
$|V_{e3}^2| \ll |V_{e1}^2|, |V_{e2}^2|$. Moreover, in the quasi-degenerate
mass range 
explored by the present experiments, 
one has $m_1^2 \gg \Delta m_{12}^2, \Delta m_{23}^2$.
Employing the above approximations
we arrive at a very simplified expression for $m_{ee}$ (see, e.g. \cite{pw}):
\be{}
|m_{ee}|^2 \approx \left[1-\sin^2 (2\theta_{12})\,
       \sin^2 \left(\frac{\phi_{12}}{2}\right)\right]\,m^2_1\,,
\label{mee2}
\ee
illustrating that \(m_{ee}\) is mostly sensitive to
\(\theta_{12}\) and \(\phi_{12}\).

\begin{figure}[t]
\centering
\includegraphics[clip,scale=0.8]{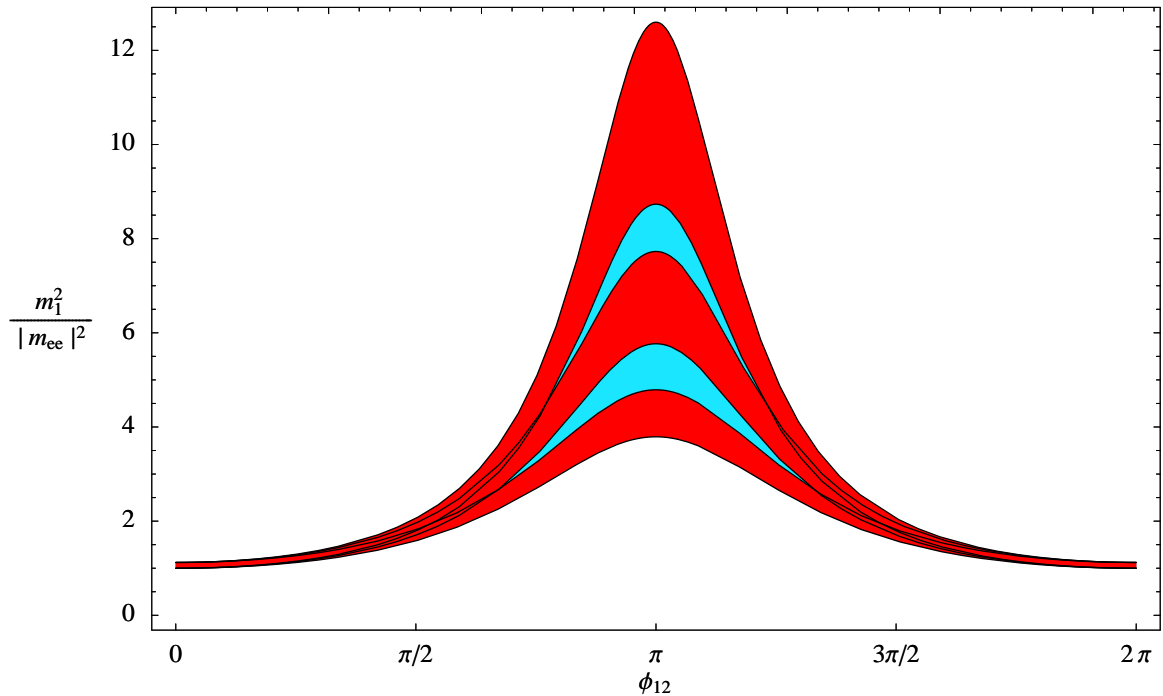}
     \caption{$m_1^2/|m_{ee}|^2$ as a function of the Majorana phase
              $\phi_{12}$.
              The full colored region is the allowed range defined by present
              2$\sigma$ limits on $\theta_{12}$, $\theta_{23}$, $\theta_{13}$
              $\Delta m^2_{12}$, $\Delta m^2_{23}$. The dark/red bands 
              correspond
              to a fixed maximal (upper band), best-fit (middle) and minimal 
              (lower)
              value for $\theta_{12}$, varying all other parameters. The second
              Majorana phase $\phi_{23}$ has been varied in the full range 
              $[0,2\pi]$.
     \label{phase}}
\end{figure}

\begin{figure}[t]
\centering
\includegraphics[clip,scale=0.8]{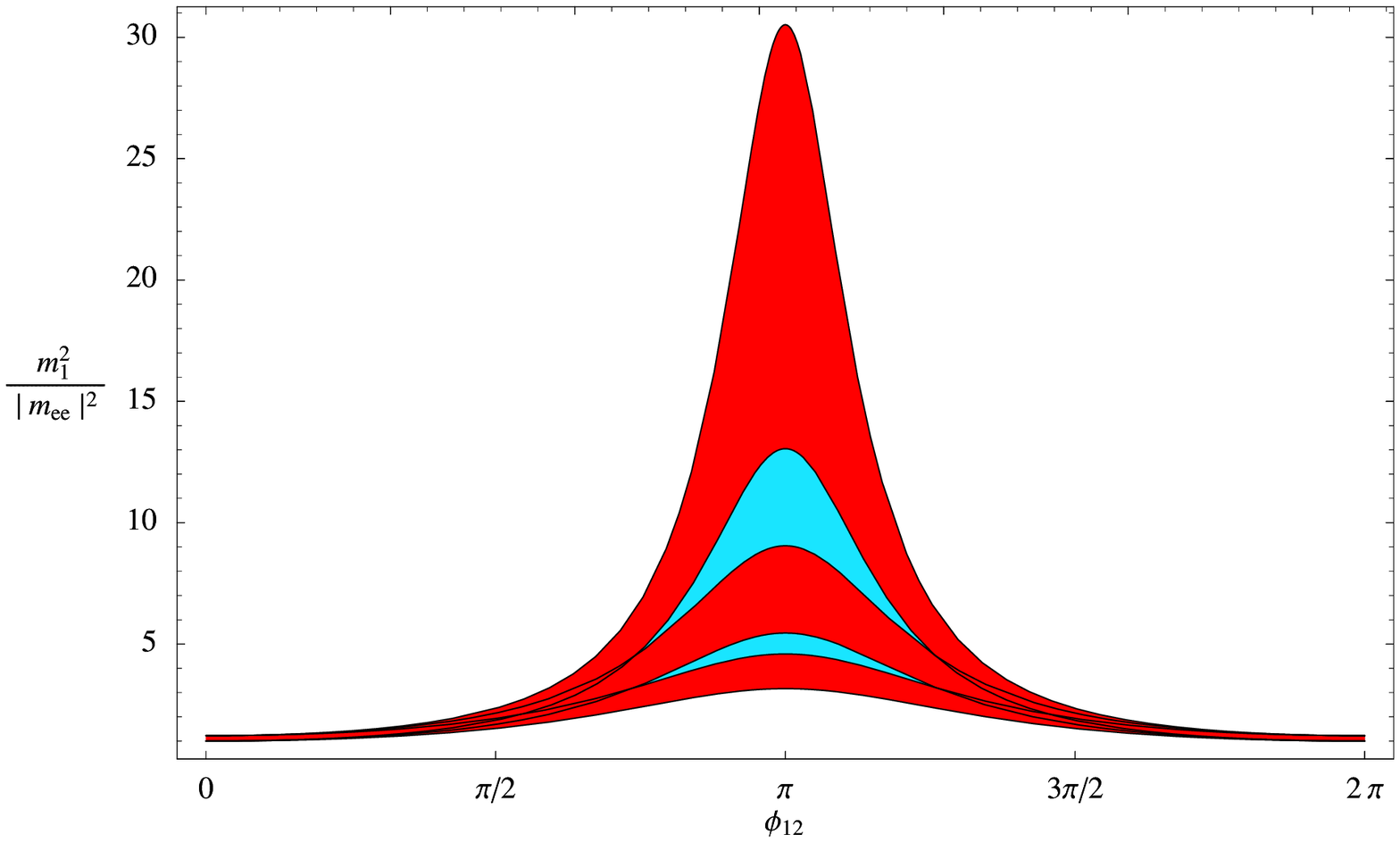}
     \caption{As Fig.~1, but with
              3$\sigma$ limits on $\theta_{12}$, $\theta_{23}$, $\theta_{13}$
              $\Delta m^2_{12}$, $\Delta m^2_{23}$.
     \label{phase2}}
\end{figure}

In Figs.~\ref{phase} and \ref{phase2}, $m_{1}^2/|m_{ee}|^2$ is
shown as a function of
\(\phi_{12}\),
using the exact relation (\ref{mee}). The full colored/shaded region
indicates the
allowed range according to the combination of $2\sigma$ (Fig.~\ref{phase})
and $3\sigma$ (Fig.~\ref{phase2})
limits  on the neutrino oscillation
observables
$\theta_{12}$, $\theta_{23}$, $\theta_{13}$, $\Delta m^2_{12}$,
$\Delta m^2_{23}$ \cite{Maltoni:2004ei}. The remaining Majorana phase
$\phi_{23}$ is varied in its full range, \(\phi_{23} \in [0,2\pi]\).
The dark/red bands correspond to a fixed maximal (upper band),
best-fit (middle) and minimal (lower) value for $\theta_{12}$, varying all
other parameters. The minimum values for $m_{1}^2/|m_{ee}|^2$ 
at $\phi_{12}=\pi$
are at 3.81 and 3.23 for $2\sigma$ and $3\sigma$ oscillation limits,
respectively.
Thus, if the experimental values for $m_{1}^2/|m_{ee}|^2$ turn out to be
smaller, a Majorana CP phase $\phi_{12}=\pi$ is excluded.
If, finally $m_{1}^2/|m_{ee}|^2<1$, the bound from unitarity of the MNS matrix
$U$ would be violated, resulting in the conclusion, that either the limit on
$m_1$ or the limit on $|m_{ee}|$ is not applicable.

Recently, a new publication of data of the Heidelberg-Moscow experiment
\cite{Klapdor-Kleingrothaus:2004wj},
searching for the double beta decay of $^{76}$Ge,
has appeared. In this work the authors have analyzed the
data taken in the period 1990-2003, and applied a new energy calibration, which
increased the previous claim for evidence \cite{HMexp}
to 4.2$\sigma$ statistical
significance.
The allowed range corresponds to \cite{dietz}
\ba{}
|m_{ee}| = \xi \cdot (0.39\textrm{ - }0.49\textrm{ eV})
~~~~~~[1\sigma], \label{mee1s}\\
|m_{ee}| = \xi \cdot (0.32\textrm{ - }0.54\textrm{ eV})
~~~~~~[2\sigma],\label{mee2s}\\
|m_{ee}| = \xi \cdot (0.24\textrm{ - }0.58\textrm{ eV})
~~~~~~[3\sigma]. \label{mee3s}
\ea
Here $\xi = {\cal M}^{SMK}/{\cal M}^{(0\nu)}$ denotes the normalization to
the NME
${\cal M}^{SMK}=4.2$, calculated in
the pn-QRPA model of \cite{smb}.
While the initial claim caused a critical debate \cite{debate}
and was not confirmed by an independent analysis
of the data \cite{HMruss}, several of these issues have been clarified in
\cite{Klapdor-Kleingrothaus:gs}. Remaining criticisms concern the exact peak
position in the energy spectrum and the relative strength of measured
background lines, which could suggest, that the observed signal is due to an
unidentified background line or that the statistical significance of the
signal is overestimated \cite{heusser}.
In any case, we feel motivated to take the evidence claim
at face value and discuss possible consequences, although we stress
that an independent test of the claimed evidence is essential, if possible with
a different double beta emitter isotope. Such a test
could be realized by the recently started
CUORICINO \cite{Arnaboldi:qy} and NEMO experiments \cite{nemo}
and the recent MPI proposal
\cite{Abt:2004yk} which revived the GENIUS proposal of the Heidelberg
group \cite{genius}.

Any conclusions about relations of the absolute neutrino mass $m_1$ and
the double beta decay observable $m_{ee}$ depend crucially on the
magnitude of the calculated NME. Typically the
uncertainty of such calculations has been estimated to be a factor 2-3,
assumed around a given central value such as the calculation in \cite{smb},
i.e. $\xi \in [0.5,2]$. In the following we will argue in favor of a
more stringent {\it upper} bound on $\xi$ and that
it constrains the allowed range of the CP Majorana phase
$\phi_{12}$. 
In Table 1 a scan of all the available matrix element calculations has
been performed. Most of the used models are based on the
proton-neutron quasiparticle random-phase approximation (pnQRPA),
like the renormalized pnQRPA, denoted by RQRPA in the table, the
self-consistent pnQRPA (SQRPA), the self-consistent RQRPA
(SRQRPA), and the fully self-consistent RQRPA (full-RQRPA). In
addition, the pnQRPA has been improved by performing a
particle-number projection on it in Ref.~\cite{Suhonen:1991db}.
This theory has been denoted by projected pnQRPA in the table. The
line with pnQRPA + pn pairing in the table denotes a theory where
proton-neutron pairing has been added to the RQRPA framework.

In these calculations various sizes of the proton and neutron
single-particle valence spaces have been used. They range from
rather modest to very extensive single-particle bases. Also, the
single-particle energies have been obtained either from the
experimental data, or, more frequently, from a Coulomb-corrected
phenomenological Woods--Saxon potential where the parameters have
been adjusted to reproduce spectroscopic properties of nuclei
close to the beta-stability line. In addition, in some
calculations the Woods--Saxon single-particle energies have been
varied close to the proton and neutron Fermi levels to reproduce
low-energy spectra of the neighboring nuclei with odd number of
protons or neutrons. Hence, remarkably diversified starting points
have been used for the calculations.

Some nuclear matrix elements of this table can be discarded 
due to various deficiencies in the theoretical frameworks
used to evaluate them. This concerns the shell-model matrix
element $\mathcal{M}^{(0\nu )}=5.00$ of Haxton and Stephenson
\cite{Haxton:1985am}, who used the weak-coupling approximation in
evaluation of it. This is quite a rough approximation, and the more
recent matrix element $\mathcal{M}^{(0\nu )}=1.74$, obtained by
performing a large-scale shell-model calculation with realistic
two-body forces, should be more reliable, although some doubts
concerning the adequacy of the size of the used single-particle
basis have been voiced. The reliability of the pnQRPA calculation
including the proton-neutron pairing has been questioned due to
the way the pairing is introduced to the theory. One can ignore
the corresponding matrix element if one wants, without changing
our final conclusions concerning the upper limit of the computed
NME. The largest matrix element was calculated
by Tomoda et al. in \cite{Tomoda:1986yz} by using a quasiparticle
mean-field based VAMPIR approach with particle-number and
angular-momentum projections included. The shortcoming of this
approach is that it does not include the proton-neutron
interaction in its framework, being
essential in description of charge-changing nuclear transitions,
which occur also in the double beta decay.

The above considerations lead to the range
\begin{equation}
      0.59 \ge \mathcal{M}^{(0\nu )} \ge 4.59
\end{equation}
for the acceptable nuclear matrix elements. Since the upper limit
of the above range comes from a pnQRPA calculation, one may ask
how does the ``$g_{\rm pp}$ problem'' of the pnQRPA affect this
value. This problem concerns the calculated matrix element of the
two-neutrino double beta decay ${\cal M}^{(2\nu)}$,
which turns out to depend strongly
on the parameter $g_{\rm pp}$, used as a scaling parameter of the
particle-particle part of the proton-neutron two-body interaction
\cite{Vogel:1986nj,Civitarese:1987px}.
However, while the uncertainty in $g_{pp}$ affects the {\it lower bound}
on NME calculations dramatically (${\cal M}^{(2\nu)}$
can even become zero for a large value of $g_{pp}$) towards lower values
of $g_{pp}$ the value for the NME enters a plateau,
making the {\it upper bound} more stable against variations in $g_{pp}$.
Moreover,
even though the
NME corresponding to the two-neutrino mode depends strongly on
$g_{\rm pp}$ within its physical range, the NME corresponding to
the neutrinoless mode depends only very weakly on this parameter.
This can be clearly seen in Figs. 1 and 2 of Ref.
\cite{Suhonen:1991db}, where the relevant double Gamow--Teller and
double Fermi matrix elements have been plotted as functions of
$g_{\rm pp}$ for the pnQRPA and projected pnQRPA calculations of
the $^{76}$Ge double beta decay. The variation of these matrix elements
around the physical value of $g_{\rm pp}\simeq 1$ is less than 20
per cent. Adding an uncertainty of 20 per cent to the above range
of acceptable values of NMEs leads us to the upper limit
\begin{equation}
      \mathcal{M}^{(0\nu )} < 5.5
\end{equation}
of the NME.

\begin{table}
  \centering
  \begin{tabular}{cll}\hline
     NME   & Theory              & References\\\hline
 5.00,1.74 & Shell model         & \cite{Haxton:1985am,Retamosa:1996rz}\\
 1.53-4.59 & pnQRPA              & \cite{smb,Suhonen:1991db,Tomoda:1990rs,Pantis:1992qe,Pantis:1996py,Aunola:1998jc,Barbero:1999tw,Suhonen:2000my,Bobyk:2000dw,Stoica:2001gh}\\
   1.50    & pnQRPA+ pn pairing  & \cite{Pantis:1996py}\\
   3.45    & projected  pnQRPA   & \cite{Suhonen:1991db}\\
   6.76    & VAMPIR              & \cite{Tomoda:1986yz}\\
 1.87-2.81 & RQRPA               & \cite{Bobyk:2000dw,Stoica:2001gh,Simkovic:1999re,Rodin:2003eb}\\
 0.59-0.65 & SRQRPA              & \cite{Bobyk:2000dw}\\
   2.40    & full-RQRPA          & \cite{Stoica:2001gh}\\
   3.21    & SQRPA               & \cite{Stoica:2001gh}\\\hline
  \end{tabular}
  \caption{Compilation of calculated nuclear matrix elements for the
neutrinoless double beta decay of $^{76}$Ge. The first column
gives the value(s) of the NME, the second
column the theory used to evaluate the NME, and the last column
the works where the quoted theory was used to evaluate the
NME.}\label{tab:MatrixElementCalculations}
\end{table}

Consequently, the limits (\ref{mee1s})-(\ref{mee3s})
read as
\ba{}
|m_{ee}|>0.30~{\rm eV}~~~~~~[1\sigma],\\
|m_{ee}|>0.24~{\rm eV}~~~~~~[2\sigma],\\
|m_{ee}|>0.18~{\rm eV}~~~~~~[3\sigma].
\ea

These {\it lower} bounds on $m_{ee}$ have to be compared to the most stringent
{\it upper} bounds on the absolute neutrino mass scale, $m_1$,
which are presently provided by data on cosmological structure formation.
According to Big Bang cosmology,
the masses of nonrelativistic neutrinos are related to the neutrino
fraction of the closure density by
$\sum_i m_i = 40\,\Omega_{\nu}\,h_{65}^2$~eV,
where $h_{65}$ is the present Hubble parameter in units of 65~km/(s~Mpc).
In the currently favored $\Lambda$CDM cosmology with a non-vanishing
cosmological constant $\Lambda$,
there is scant room left for the neutrino component.
The free-streaming relativistic
neutrinos suppress the growth of fluctuations
on scales below the horizon
(approximately the Hubble size $c/H(z)$)
until they become nonrelativistic at redshifts
$z\sim m_j/3T_0 \sim 1000\,(m_j/{\rm eV})$.

\begin{figure}[t]
\centering
\includegraphics[clip,scale=0.8]{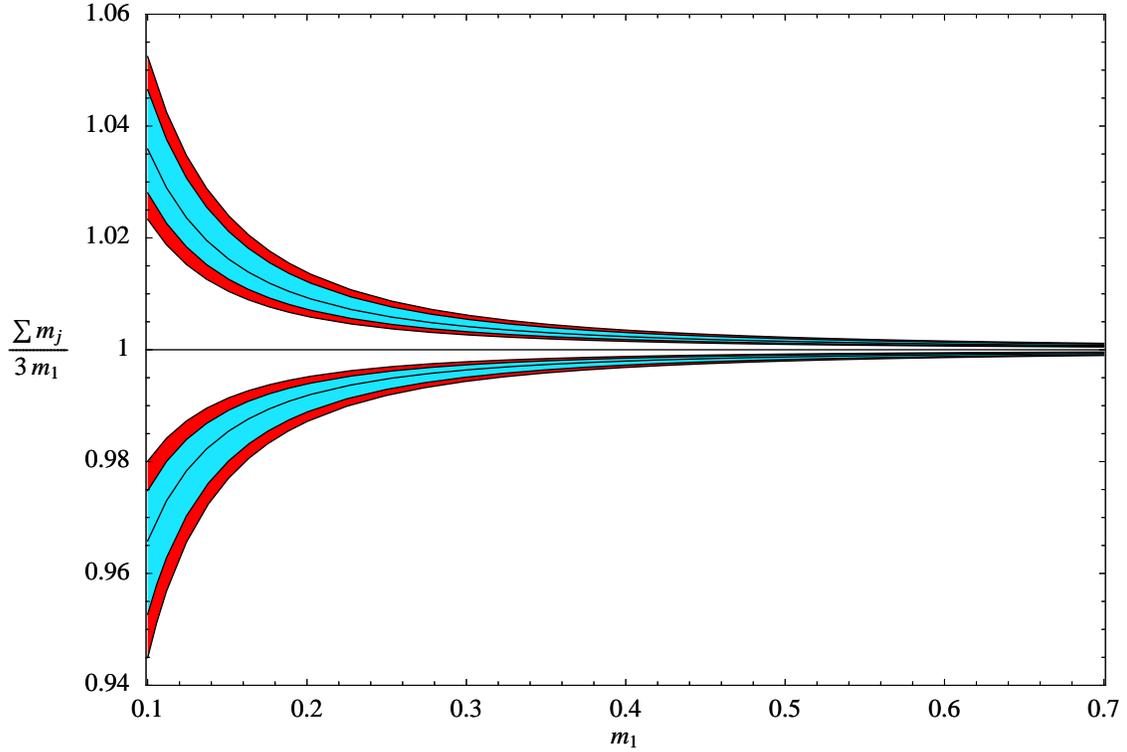}
     \caption{Relation of  the cosmologically relevant sum of neutrino masses,
     $\sum_i m_i$ and the lightest neutrino mass $m_1$. The upper and lower
     branches correspond to normal and inverse hierarchy, respectively.
     The light/blue and dark/red bands indicate the $2\sigma$ and $3\sigma$
     range of $\Delta m^2_{12}$ and $\Delta m^2_{23}$, while the central line
     marks the best fit.
     \label{sumnu}}
\end{figure}

Recent limits, obtained by combining CMB measurements with data on the
large scale structure of the universe, imply
an upper bound on the sum of the three neutrino mass eigenstates
\cite{Hannestad:2004bu},
\be{}
\sum_i m_i < 0.42\textrm{ - } 1.80\textrm{ eV}~~~~~~[2\sigma].
\ee
The exact cosmological bound depends on the data and specific priors
used in the analysis:
The weakest bound utilizes only data from WMAP and the Sloan Digital Sky
Survey. In particular, limits utilizing the Lyman-$\alpha$ forest, the
absorption observed in quasar spectra by neutral hydrogen in the
intergalactic medium, provide more stringent bounds \cite{wlya}
\be{}
\sum_i m_i < 0.42\textrm{ - } 0.69\textrm{ eV}~~~~~~[2\sigma],
\ee
as
compared to the data sets without the Lyman-$\alpha$ forest \cite{wolya},
\be{}
\sum_i m_i < 0.6\textrm{ - } 1.8\textrm{ eV}~~~~~~[2\sigma].
\ee
Since the effects of possible systematics in this data set need still to be
explored further, both observationally and theoretically, each set 
of analyses
should be discussed here.
Within the present decade, the combination of SDSS data with CMB data of the
PLANCK satellite will obtain a $2~\sigma$ detection threshold
on $\sum_i m_i$ close to 0.1-0.2~eV \cite{Hannestad:2004bu,futurelimits}.
Fig.~\ref{sumnu} shows the relation of the sum of neutrino masses with $m_1$,
\be{}
\sum_i m_i = m_1 \left(1 + \sqrt{1+\frac{\Delta m^2_{12}}{m_1^2}}
+ \sqrt{1+\frac{\Delta m^2_{12}}{m_1^2}\pm\frac{\Delta m^2_{23}}{m_1^2}}
\right),
\ee
for normal and inverse hierarchy in the upper and lower panel, respectively.
The different curves indicate the best-fit and upper and lower
$2\sigma$ and $3\sigma$ ranges.

Combining the resulting bound on $m_1$ with the range for the effective mass
$m_{ee}$ given in \cite{Klapdor-Kleingrothaus:2004wj}, one
obtains:
\be{}
m_1^2/|m_{ee}|^2< 0.36-0.93~(0.64-1.7)~~~~\textrm{at}~~~~2\sigma~(3\sigma)
\ee
for data sets using the Ly-$\alpha$ forest and
\be{}
m_1^2/|m_{ee}|^2< 0.71-6.3~(1.3-11.1)~~~~\textrm{at}~~~~2\sigma~(3\sigma)
\ee
without the Ly-$\alpha$ forest.
It is obvious, that the $2\sigma$ range for the double beta decay observable
is in conflict with all cosmological fits including the Ly-$\alpha$ forest and
with some without the Ly-$\alpha$ forest. This may indicate that either
the double beta decay signal is due to a statistical fluctuation or due to a
mechanism involving exchange of other particles besides light massive Majorana
neutrinos. Examples for the latter case include, e.g. sparticles in
$R$-parity violating supersymmetry, leptoquarks, or right-handed neutrinos
and $W$ bosons (for an overview see \cite{npznbb}).
Alternatively, the cosmological neutrino mass bound may not be applicable,
e.g. by introducing broken scale-invariance in the primordial power spectrum
\cite{Blanchard:2003du},
or due to fast decays of the relic neutrino background \cite{Beacom:2004yd}.
The $3\sigma$ range for $m_{ee}$, however, is still compatible with most of
the cosmological bounds although in most cases it is smaller than the
minimum value, $m_{1}^2/|m_{ee}|^2=3.81$ for $\phi_{12}=\pi$.
If a value  $1<m_{1}^2/|m_{ee}|^2<3.81$ will be confirmed in future
cosmological data fits or in the upcoming tritium beta decay
spectrometer KATRIN
\cite{katrin}, Majorana CP phases around $\phi_{12}=\pi$ can be excluded.

In conclusion, we discussed upper bounds on nuclear matrix elements
and their implications for the Majorana CP phase $\phi_{12}$, when
the recent evidence claim for neutrinoless double beta
decay and the cosmological neutrino mass
bound are combined.
We deduced that for a combination of  $2\sigma$ experimental limits in most
analyses the mass mechanism interpretation of the double beta decay evidence
is incompatible with the cosmological neutrino mass bound.
On the other hand,
the range of the Majorana phase
$\phi_{12}$ can be constrained, when combining $3\sigma$ experimental limits
with reasonable upper bounds for the nuclear matrix elements.
Assuming CP conservation, the CP phase factors $\exp(i \phi_{ij})$
are reduced to CP parities $\eta_{ij}=\pm 1$. In this case the neutrino
CP parity is fixed to $\eta_{12}= +1$.

\section*{Acknowledgements}
This work was supported by the Bundesministerium f\"ur
Bildung und Forschung (BMBF, Bonn, Germany) under the contract number
05HT1WWA2.

\clearpage

\end{document}